\documentclass{eas}
\usepackage{graphicx}
\TitreGlobal{Physics of Evolved Stars 2015}
\begin{document}

%%-----------------------------
%%      the top matter
%%-----------------------------
\title{The role of binarity in Wolf-Rayet central stars of planetary nebulae} 
\runningtitle{The role of binarity in [WR] CSPNe}
\author{Brent Miszalski}
\address{South African Astronomical Observatory, PO Box 9, Observatory,
7935, South Africa}\secondaddress{Southern African Large Telescope Foundation, PO Box 9,
Observatory, 7935, South Africa}
\author{Rajeev Manick}\address{Instituut voor Sterrenkunde, KU Leuven, Celestijnenlaan
  200D bus 2401, B-3001 Leuven, Belgium}
\author{Vanessa McBride}\address{Astrophysics, Cosmology and Gravity Centre, Department of
Astronomy, University of Cape Town, Private Bag X3, Rondebosch 7701, South Africa}\secondaddress{South African Astronomical Observatory, PO Box 9, Observatory, 7935, South Africa}
\begin{abstract}
   Over a hundred planetary nebulae (PNe) are known to have H-deficient central stars that mimic the spectroscopic appearance of massive Wolf-Rayet stars. The formation of these low-mass Wolf-Rayet stars, denoted [WR] stars, remains poorly understood. While several binary formation scenarios have been proposed, there are too few [WR] binaries known to determine their feasibility. Out of nearly 50 post-common-envelope (post-CE) binary central stars known, only PN~G222.8$-$04.2 ([WC7], $P=1.26$ d) and NGC~5189 ([WO1], $P=4.05$ d) have a [WR] component. The available data suggests that post-CE central stars with [WR] components lack main sequence companions and have a wider orbital separation than typical post-CE binaries. There is also some indirect evidence for wide binaries that could potentially lead to the discovery of more [WR] binaries. 
\end{abstract}
\maketitle
%%-----------------------------
%%      your text
%%-----------------------------
\section{Post-common-envelope binaries}
Over the last few years the number of planetary nebulae (PNe) known to have a binary central star has increased dramatically (Miszalski et al. 2009a, 2011a; Jones 2015). Most of these have orbital periods of 0.1--1.0 d, meaning they have passed through the poorly understood common-envelope (CE) phase (Ivanova et al. 2013). Post-CE PNe are the youngest post-CE population ($\sim$10$^4$ yr) and are therefore a key strategic population with which to study the CE interaction (e.g. Boffin \& Miszalski 2011; Tocknell et al. 2014; Jones 2015). The vast majority of post-CE central stars of PNe (CSPNe) have white dwarf (WD) primaries with main sequence (MS) secondaries, but double WD (double degenerate, DD) systems are also common. Boffin et al. (2012) demonstrated the importance of radial velocity (RV) monitoring to discover DD systems in which there is no sign of a binary in optical lightcurves. There is therefore a potentially large population of PNe with evolved WD companions whose binaries cannot be found via photometric monitoring.

We have commenced an RV monitoring survey of one such population (Manick et al. 2015). These are the Wolf-Rayet CSPNe that appear in over 100 PNe (see Todt \& Hamann 2015 for a review), low-mass analogues of massive WR stars (denoted [WR] stars). The aim of this work is to identify post-CE binaries with one component being a [WR] star to improve our understanding of the origin of [WR] stars. No such systematic RV monitoring of [WR] CSPNe has been made in the past, perhaps because of concerns that the strong winds would dilute any binary signature. Despite the apparent hurdles involved, we were grateful to have had the encouragement of Olivier Chesneau to pursue this project during his visit to Cape Town in early 2013. RV monitoring and in particular, the cross-correlation technique, is a successful and routine means to find massive WR binaries (e.g. Foellmi et al. 2003). We are applying the same methodology as Foellmi et al. (2003) to high quality multi-epoch spectra (2 \AA\ resolution and signal-to-noise $>$ 40 in the continuum).

While the atmospheric composition of most [WR] stars belonging to the carbon sequence ([WC] types) can be readily explained (Werner \& Herwig 2006; Todt \& Hamann 2015), those of the nitrogen sequence ([WN] types) cannot. Their extremely He-rich atmospheres ($>$90\% by mass, see Miszalski et al. 2012a and Todt et al. 2013) suggests that [WN] stars may form via more peculiar channels, e.g. a binary merger (Miszalski et al. 2012a). Several other binary formation scenarios have also been proposed for [WR] stars (e.g. De Marco \& Soker 2002). Identifying post-CE [WR] binaries will not only provide distance-independent stellar parameters of [WR] stars, but will also help determine whether such binary scenarios can feasibly explain [WR] stars.

At present only two post-CE [WR] binaries are known out of nearly 50 post-CE PNe (Jones 2015). Hajduk et al. (2010) found periodic photometric variability in the [WC7] nucleus of PN~G222.8$-$04.2 with an orbital period of 0.63 or 1.26 d (the latter is the correct period, see Miszalski et al. 2015). Manick et al. (2015) used the Southern African Large Telescope to discover a 4.04 d periodic variability in the [WO1] nucleus of NGC~5189. Miszalski et al. (2015) presented an updated RV curve for NGC~5189 with a revised 4.05 d orbital period. Some other [WR] CSPNe monitored by Manick et al. (2015) were not found to be variable and several more objects are also being monitored in this ongoing project (see also Miszalski et al. 2015). The mass functions derived from the RV curves suggest an evolved WD companion for NGC~5189 ($M_2>0.6$ $M_\odot$) and perhaps a subdwarf O/B-type companion for PN~G222.8$-$04.2 ($M_2\sim1.6$ $M_\odot$). Further work is required to constrain the orbital inclination of both systems.

The nebular morphology of NGC~5189 (Sabin et al. 2012) is typical of post-CE PNe with a high degree of low-ionisation filaments and collimated outflows (Miszalski et al. 2009b). It bears particular resemblance to the post-CE PN NGC~6326 (Miszalski et al. 2011b). The 4.05 d orbital period is unusually long for a post-CE binary and suggests that post-CE [WR] binaries may need longer orbital periods to accommodate the extended atmosphere of the [WR] component. The lack of strong irradiation effects in several [WR] and other H-deficient CSPNe (Ciardullo \& Bond 1996; Gonz{\'a}lez P{\'e}rez et al. 2006), combined with the WD nature of the companion in NGC~5189, suggests other [WR] CSPNe may also have WD companions that can only be found via RV monitoring.

\section{Wide binaries}
Much less is known about binary CSPNe at wider orbital separations (several months to years). Van Winckel et al. (2014) presented the first RV observations for wide CSPNe, namely a 1105 d orbital period in PN G052.7$+$50.7 and a very long orbital period in LoTr~5. The latter is one of four barium CSPNe in which an s-process and/or carbon-enhanced red giant secondary dominates at optical wavelengths, while an evolved WD can only be seen at UV wavelengths. The fast rotating red giant accreted the polluted material from the progenitor of the WD earlier in its evolution. RV monitoring of the remaining barium CSPNe, WeBo~1 (Bond et al. 2003), A~70 (Miszalski et al. 2012b) and Hen~2-39 (Miszalski et al. 2013), should also reveal long term RV orbits. 

In terms of [WR] CSPNe, the best candidate for a wide binary is the poly-polar PN CPD-56$^\circ$ 8032 (Hen~3-1333, Chesneau et al. 2006). It has a resolved dust disk with an inner radius of 97 AU and a binary may be necessary to explain its presence in Hen~3-1333 (Van Winckel et al. 2009). Hen~3-1333 also features dual-dust chemistry that is often associated with binarity. An intriguing feature of Hen~3-1333 is that the central star experiences quasi-periodic dust obscuration events every $\sim$5 yrs (see Miszalski et al. 2011c), possibly the result of a periastron interaction with a companion. Miszalski et al. (2011c) could reproduce the dust formation at 97 AU, much further out than where a companion in a 5 yr orbit would be located. Manick et al. (2015) rule out a close binary based on RV measurements. We encourage further observations of this intriguing object to determine whether a wide binary is present. Another promising candidate is Hen~2-113, which has several features in common with Hen~3-1333 (Lagadec et al. 2006).

%%-----------------------------
%%      your bibliography
%%-----------------------------

\end{document}